# Domain-Selective Enhancement of Second Harmonic Generation in Monolayer $MoS_2$ via Ferroelectricity-Controlled Photodoping

*David Hernández-Pinilla,*[1,2] *Line Jelver,*[3] *María Jesús Martínez-Morillo,*[1,2] *César Hernando-Fuente,*[1] *Miquel Cherta,*[1] *Guillermo López-Polin,*[1,4] *Joel D. Cox,*[3,5] *Luisa E. Bausá,*[1,2,4] *and Mariola O Ramírez*[1,2,4,*]

1. Dept. Física de Materiales, Universidad Autónoma de Madrid, Madrid 28049, Spain
2. Instituto Nicolás Cabrera, Universidad Autónoma de Madrid, Madrid 28049, Spain
3. POLIMA - Center for Polariton-driven Light−Matter Interactions, University of Southern Denmark, DK-5230 Odense M, Denmark
4. Condensed Matter Physics Center (IFIMAC), Universidad Autónoma de Madrid, Madrid 28049, Spain
5. Danish Institute for Advanced Study, University of Southern Denmark, DK-5230 Odense M, Denmark

*Corresponding autor: mariola.ramirez@uam.es

**Abstract text.** Hybrid heterostructures combining two-dimensional semiconductors with ferroelectric materials offer a versatile route to actively control light–matter interactions at the nanoscale. Here, we report all-optical, light-induced domain-selective control of second-harmonic generation (SHG) in monolayer molybdenum disulfide ($MoS_2$) integrated with periodically poled lithium niobate ($LiNbO_3$). Spatially resolved SHG imaging reveals a pronounced modulation of the nonlinear optical response of monolayer $MoS_2$ governed by the ferroelectric domain pattern of the underlying substrate. A strong SHG contrast is observed between domains of opposite polarization, with a marked dependence on both the excitation wavelength and the incident optical power. The comparison between ferroelectric domains that either enable or do not exhibit light-driven photodoping in the $MoS_2$ monolayer provides a direct assessment of the role of carrier density in the nonlinear optical response. We find that ferroelectric-polarization-controlled photodoping at the $MoS_2$/$LiNbO_3$interface enhances the effective second-order susceptibility, $\chi^{(2)}$, producing an increase in SHG intensity of up to ~70% under resonant excitation conditions. Ab initio calculations corroborate that charge doping modifies the electronic band structure of $MoS_2$ and strongly affects $\chi^{(2)}$ in the resonant regime, providing microscopic support for the experimentally observed modulation. The results highlight the combination of light intensity and ferroelectricity as a powerful knob for band-structure modulation in 2D materials and reconfigurable nonlinear optical responses, opening pathways toward programmable frequency conversion, smart light modulators, and advanced nonlinear photonic functionalities in integrated hybrid platforms.

## 1. Introduction

Transition metal dichalcogenides (TMDs) are an attractive class of semiconducting materials with exceptional optical properties, particularly in their two-dimensional (2D) monolayer form, where they exhibit a direct bandgap and strong excitonic effects.[1-4] The pronounced light–matter interactions in TMDs support a wide range of linear and nonlinear optical responses. Among the latter, second-order nonlinear processes such as second-harmonic generation (SHG) occur in monolayers and odd-layered structures as a consequence of inversion-symmetry breaking, enabling a nonzero second-order susceptibility, which can become particularly large under resonant conditions involving excitonic or interband transitions.[5-8]

In addition, a key feature of 2D TMDs is the possibility to dynamically tune their optical response by modifying the carrier density via external stimuli such as electrostatic gating, chemical doping, or strain engineering.[9-12] This tunability enables reversible control over excitonic resonances, significantly modifying their absorption and emission characteristics. For instance, gate-tunable photoluminescence (PL) in monolayer $MoS_2$ has been demonstrated through active switching between neutral and charged excitons (trions), opening avenues for excitonic modulators.[13,14] Carrier injection has also been shown to modulate the nonlinear optical susceptibility under resonant conditions, enabling electrical control over both second- and third-order nonlinear responses in gated devices.[15-18]

An alternative route to modulate charge carrier doping in TMDs is provided by ferroelectric substrates, which offer non-volatile, reversible, and spatially resolved control of their optical and electronic properties.[19-23] Recent reports have demonstrated the creation of reconfigurable p-n junctions in $MoS_2$ coupled to periodically poled $LiNbO_3$, achieved through either domain-selective photodoping[24] or pyroelectric charges generated by cooling the system to cryogenic temperatures.[25] Later studies have extended this approach to $WSe_2$, where p-n photodetectors integrated with periodically poled $LiNbO_3$ or even the ferroelectric modulation of quantum emitters have been recently reported.[26-28] Together, these studies highlight the potential of ferroelectric domain engineering as an active platform for extending the functionalities of 2D materials and enabling new reconfigurable devices within nanophotonics and nanoelectronics.

To date, however, the effect of ferroelectricity on the nonlinear optical response of 2D TMDs has been scarcely investigated. Only a few studies have reported that interfacial polar symmetry coupling in heterostructures of $MoS_2$ and lead zirconate titanate (PZT) can modulate the SHG response, altering its intensity and polarization patterns.[29,30] Yet the broader influence of ferroelectric substrates on the nonlinear optical properties of 2D materials remains largely unexplored. In addition to polar ordering and non-volatile out-of-plane electric fields, ferroelectrics exhibit a variety of functional properties that can be exploited for the design of novel ultrathin nonlinear devices with actively tunable functionalities. The strong interaction of ferroelectric materials with light has led to a variety of all-optical phenomena, including transient and permanent polarization reversal, domain wall motion, or bulk photovoltaic effects.[31-35] Nonetheless in hybrid 2D–ferroelectric systems, this capability has so far been harnessed only for optoelectronic and neuromorphic applications.[36-38]

In this work, we exploit the combined effect of light and ferroelectricity to dynamically control the nonlinear optical properties (SHG) of monolayer $MoS_2$ (1L-$MoS_2$) deposited on the polar surface of a periodically poled $LiNbO_3$ (PPLN) substrate. Given the relevance of these materials in photonics and optoelectronics, the SHG performance of $MoS_2$ and $LiNbO_3$ has been extensively investigated, albeit independently. In monolayer $MoS_2$, a variety of frequency conversion processes, such as optical parametric amplification,[39] electrical control of the harmonic generation efficiency,[16] and optical modulation of the polarization and amplitude of SHG signals[40] have been recently reported. On the other hand, periodically poled $LiNbO_3$ (PPLN) has long served as a prototypical ferroelectric material for nonlinear optics and communication technologies. Indeed, recent developments in thin-film $LiNbO_3$ have significantly expanded its use in integrated photonic circuits and devices.[41,42] The integration of monolayer $MoS_2$ with PPLN thus constitutes a promising approach for the development of hybrid nonlinear optical systems and next-generation optoelectronic devices. By studying SHG in this integrated platform, we demonstrate how ferroelectric domain engineering and light can cooperatively tune the quadratic nonlinear response of a 2D material in a spatially selective manner.

Specifically, we demonstrate a light-induced ferroelectric modulation of the SHG response in $MoS_2$, which depends on the ferroelectric domain orientation of the underlying substrate and enables spatial tuning of the quadratic nonlinear response of the 2D material. This modulation yields a relative spatial difference of up to 70% in the SHG intensity between 1L-$MoS_2$ on domains with opposite polarity. A key advantage of this ferroelectric platform is that domains with opposite polarization create two local environments within the same sample: one in which light-induced photodoping occurs in the $MoS_2$ monolayer and another in which photodoping remains negligible. This built-in comparison enables us to directly assess the role of carrier density in the nonlinear optical response of $MoS_2$ and identify ferroelectricity-controlled photodoping as the origin of the enhanced SHG signal. Indeed, the spatial modulation of the SHG response originates from a domain-selective, photoinduced charge generation and transfer process at the $MoS_2$/PPLN interface, which increases the electron density in domains with upward spontaneous polarization and enhances the SHG signal, in agreement with ab initio calculations.

On this basis, we systematically study the ferroelectric-domain-dependent modulation of the SHG response as a function of excitation wavelength and intensity, light polarization, and crystal orientation, highlighting the combined role of optical excitation and ferroelectricity. The demonstrated enhancement and dynamic control of SHG in low-dimensional materials establish a robust all-optical route to engineer nonlinear light–matter interactions at the nanoscale, with potential applications in a broad range of photonic technologies, including quantum optics, nonlinear spectroscopy, biophysics, and integrated photonics.

## 2. Results and discussion

**Figure 1a** shows an optical micrograph of a large-area 1L-$MoS_2$ deposited on the polar surface of a (Z-cut) PPLN substrate. The monolayer was transferred in such a way that its armchair direction (y-axis of the $MoS_2$ crystal lattice) is parallel to the ferroelectric domain walls of the $LiNbO_3$ (y-axis of $LiNbO_3$). For SHG experiments, a reflection geometry was employed (Figure 1b). The spatial distribution of the SHG intensity recorded at normal incidence is shown in Figure 1c. As observed, it reveals a clear spatial modulation of the 1L-$MoS_2$ SHG intensity that correlates with the orientation of the underlying spontaneous polarization of the substrate, which alternates between up and down ferroelectric domains. Such modulation is not observed in the SHG originating from the bare $LiNbO_3$, which is independent of the ferroelectric domain orientation, as shown at the bottom of the image. Here it is worth mentioning that the SHG signal from $MoS_2$ is generated at around 400 nm wavelength, spectrally overlapping the C interband transition of 1L-$MoS_2$ (**Figure S1**). At this wavelength, the SHG intensity from the monolayer is approximately two orders of magnitude higher than that from bare $LiNbO_3$ (**Figure 1d** and **S2**). Figure 1e provides a quantitative comparison of the spatial intensity profiles recorded from both regions, 1L-$MoS_2$ and bare PPLN. The results show that 1L-$MoS_2$ not only exhibits a much larger intensity but also displays a spatial modulation depth of approximately 70% between domains with upward and downward polarization (hereafter denoted as $P_{up}$ and $P_{down}$, respectively) with maximum values observed on $P_{up}$ domains. In contrast, the PPLN substrate exhibits a nearly constant SHG intensity across domain surfaces with opposite polarities. The only noticeable changes occur at the ferroelectric domain walls, where a decrease in the SHG signal is observed, in agreement with previous reports.[43]

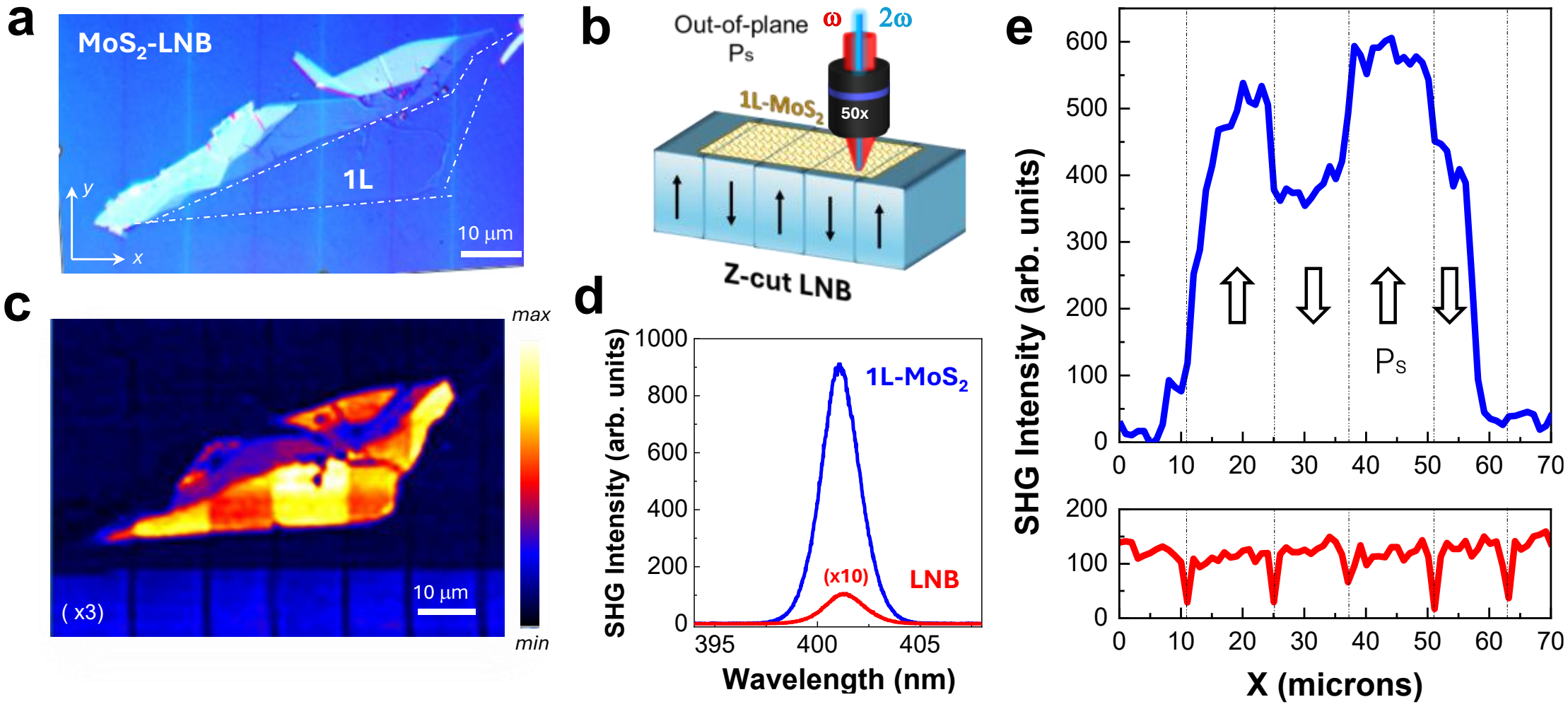

**Figure 1. Spatial modulation of SHG in 1L-$MoS_2$ on PPLN.** a) Optical micrograph of a large-area 1L-$MoS_2$ transferred onto a *z*-cut (polar) surface of PPLN substrate, with the armchair direction of the 1L-$MoS_2$ (y-axis) aligned parallel to the ferroelectric domain walls ($LiNbO_3$ y-axis). The 1L-$MoS_2$ region is indicated by the dashed line. b) Schematic illustration of the 1L-$MoS_2$/$LiNbO_3$ heterostructure used in SHG experiments. The spontaneous polarization (Ps) of $LiNbO_3$ is represented by black arrows. Excitation (red) and SHG (blue) signals are collected in a reflectance configuration. c) Spatial distribution of SHG intensity recorded at normal incidence using a fundamental wavelength of 800 nm and an incident power of 17 mW. Lower part: SHG from the bare $LiNbO_3$ substrate, recorded with an excitation power increased by a factor of 3 to enable comparison with 1L-$MoS_2$ within the same scan. d) SHG spectra from 1L-$MoS_2$ (blue) and bare $LiNbO_3$ (red) measured for a fundamental wavelength of 800 nm; the $LiNbO_3$ signal was multiplied by a factor of 10 for comparison. e) Spatial profiles of the integrated SHG intensity for the $MoS_2/LiNbO_3$ heterostructure (blue) and bare $LiNbO_3$ (red) measured along the horizontal direction across several ferroelectric domains (indicated by the vertical dashed lines).

**Figure 2a** displays representative SHG maps obtained at different fundamental wavelengths between 780 and 900 nm, while keeping the fundamental beam power constant. Within this spectral range, the SHG intensity of 1L-$MoS_2$ is markedly enhanced on both $P_{up}$ and $P_{down}$ domains, reaching maximum values near 840 nm due to the two-photon resonance with the C interband transition of 1L-$MoS_2$, located at 440 nm. Figure 2b compares the SHG spatial profiles across the PPLN substrate at three representative wavelengths: on-resonance (840 nm) and off-resonance at the lower (780 nm) and upper (900 nm) ends of the tuning range. Both the absolute SHG intensity and the modulation depth, defined as $(I_{up}-I_{down})/I_{down}$, exhibit a clear dependence on the excitation wavelength. Namely, at the resonant wavelength, 1L-$MoS_2$ exhibits a pronounced enhancement of the SHG intensity, and a maximal SHG contrast on domains with opposite polarization orientations. Conversely, at the spectral edges (e.g., 780 nm or 900 nm), the SHG signal is significantly reduced, and the domain-dependent modulation depth decreases. The wavelength-dependent spatial modulation of the SHG response is displayed in Figure 2c. As observed, the modulation depth closely follows the C transition, indicating that the resonant enhancement of $\chi^{(2)}$ in $MoS_2$ strengthens the ferroelectric modulation of the SHG intensity in 1L-$MoS_2$-PPLN.

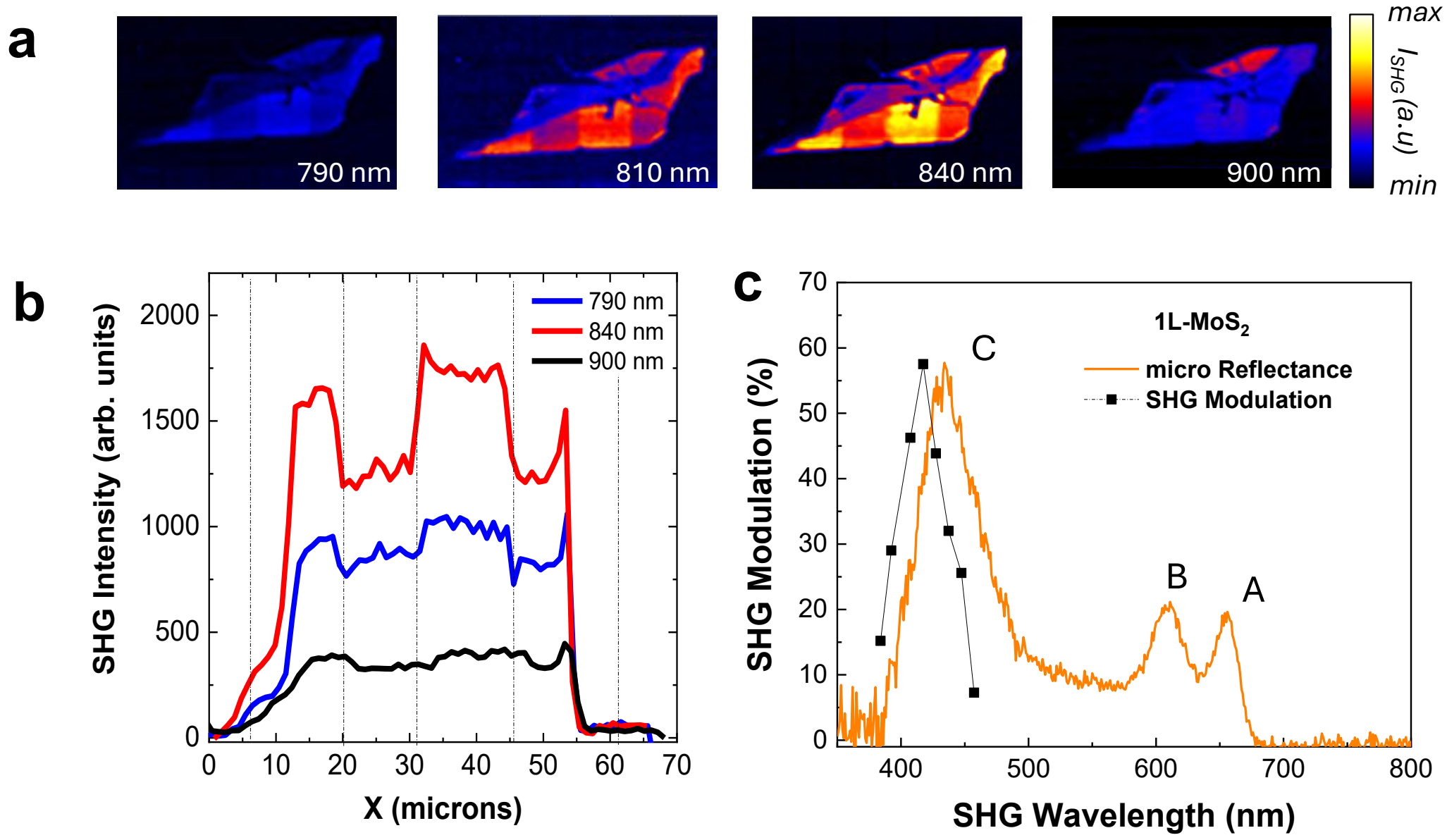

**Figure 2. Spectral tuning of SHG modulation.** a) Representative SHG intensity maps recorded at fundamental wavelengths in the 790-900 nm spectral range, using an excitation power of 20 mW. b) Spatial

profiles of the integrated SHG intensity for the 1L-$MoS_2$/$LiNbO_3$ heterostructure measured across domain surfaces with opposite polarities at fundamental wavelengths of 790 nm (blue), 840 nm (red) and 900 nm (black). c) Fundamental- wavelength dependence of the SHG modulation between $P_{up}$ and $P_{down}$ domains (black) together with the differential reflectance spectrum of 1L-$MoS_2$ (orange) for comparison with its C-band resonance.

To determine the origin of the SHG modulation, we next study the influence of light polarization on the SHG patterns. Here it is important to note that 1L-$MoS_2$ and $LiNbO_3$ share the same angular polarization dependence in their SHG signals. This behaviour originates from the symmetry of the $\chi^{(2)}$ tensor, which is determined by their respective point groups ($D_{3h}$ for 1L-$MoS_2$ and $C_{3v}$ for $LiNbO_3$). Thus, by carefully aligning the crystallographic axes of both systems and choosing appropriate polarization configurations, it is possible to combine the SHG signals from both materials or selectively suppress one contribution without affecting the other. This allows us to rule out the possibility of interfacial polar symmetry coupling previously reported for other different systems.[15,16] **Figure 3** shows the effect of the input-output polarization on the SHG spatial distribution for two monolayers: one with its armchair direction parallel to the $LiNbO_3$ domain walls (Fig. 3a) and another oriented perpendicular to them (Fig. 3b). In the first case, when the *y* axes of $MoS_2$ and $LiNbO_3$ are aligned (Fig. 3a), both systems exhibit maxima and minima under the same polarization conditions. In contrast, a 90° rotation of 1L-$MoS_2$ with respect to $LiNbO_3$ (Fig. 3b) decouples the two systems, leading to opposite behavior. That is, the maximum SHG intensity from $MoS_2$ corresponds to a minimum (or zero) intensity from $LiNbO_3$ and vice versa, consistent with their angular polarization-dependent properties (see **Figure S3** for additional polarization plots).

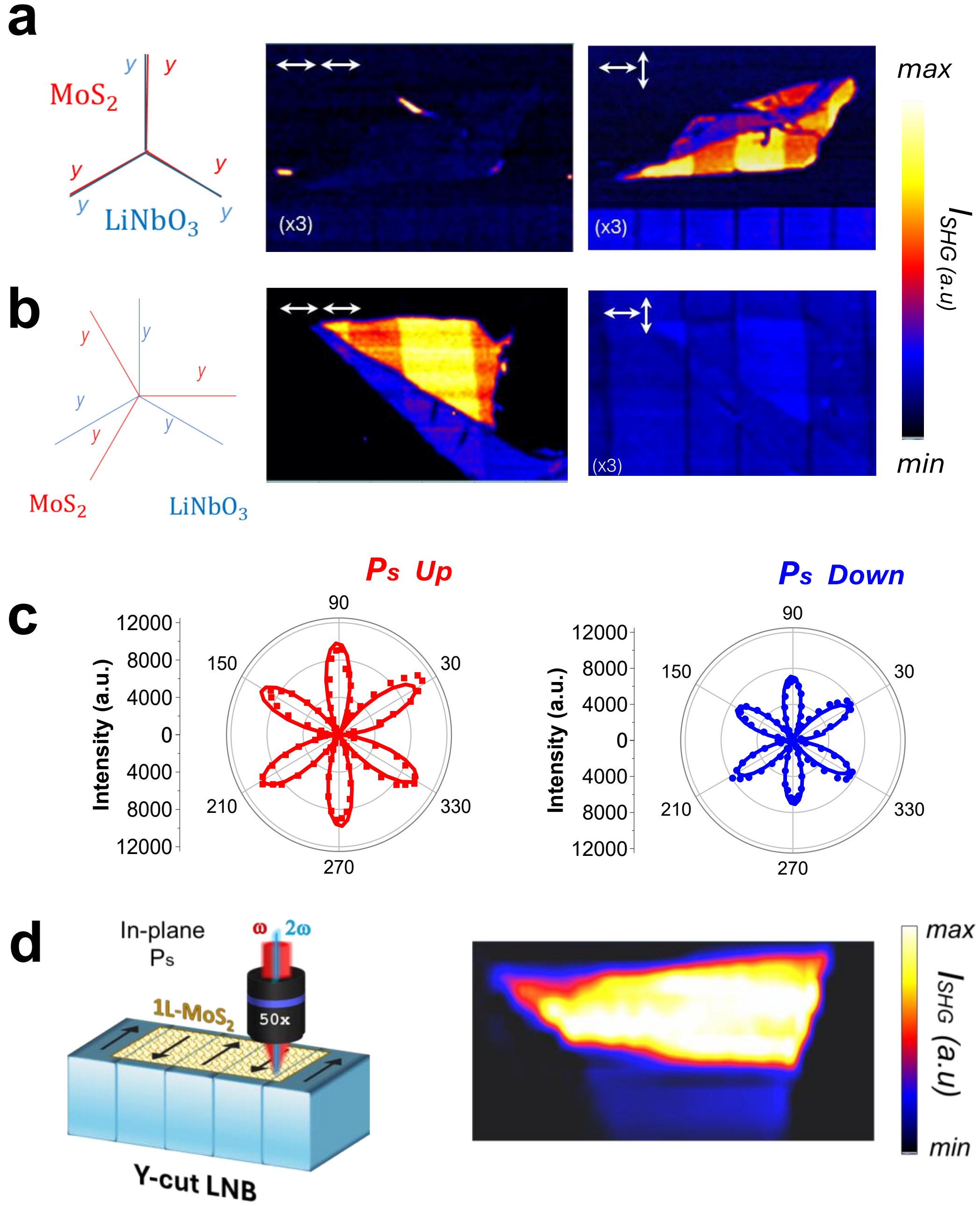

**Figure 3. Polarization dependence of SHG modulation.** Spatial distribution of the integrated SHG intensity recorded at normal incidence for orthogonal input-output light polarization configurations in two different flakes, with the 1L-$MoS_2$ armchair direction oriented (a) parallel and (b) perpendicular to the ferroelectric domain walls of $LiNbO_3$. Arrows indicate the linear polarization state of the fundamental and the corresponding SHG signal, respectively; the crystallographic axes are also indicated. c) Polar plots of the SHG signal from 1L-$MoS_2$ recorded at normal incidence on $P_{up}$ (red) and $P_{down}$ (blue) ferroelectric domains. d) Left: Schematic illustration of the $MoS_2$ / Y-cut $LiNbO_3$ heterostructure. The spontaneous polarization ($P_s$) of $LiNbO_3$ is represented by black arrows, indicating the in-plane (nonpolar) configuration. Right: SHG intensity map of 1L-$MoS_2$ flakes transferred onto the nonpolar surface of $LiNbO_3$.

This opposite behavior excludes a possible coherent superposition of the SHG fields from the $LiNbO_3$ substrate and the 1L-$MoS_2$. Additionally, the modulation depth of the $MoS_2$ SHG signal between oppositely polarized domains is nearly identical for both crystallographic orientations (Fig. 3a,b) and comparable to that measured under unpolarized conditions (without an analyzer). Furthermore, across all polarization configurations (Fig. 3c), the up and down domains preserve the same angular dependence, differing only in their overall intensity. These observations consistently rule out interfacial polar coupling or interference effects as the dominant mechanism underlying the modulation.At this stage it is useful to compare these results with those obtained on a nonpolar surface of the substrate. Figure 3d shows a representative SHG map obtained from a 1L-$MoS_2$ transferred onto the nonpolar Y-cut surface of the PPLN crystal. As observed, the SHG from the 1L-$MoS_2$ remains spatially uniform across the surface, with no detectable modulation, in sharp contrast to the behavior on the polar surface.

Overall, these results suggest that the observed SHG modulation can be attributed to a modulation of the electron doping in $MoS_2$ on the polar surface of PPLN, which alters the second-order susceptibility, $\chi^{(2)}$, near the C-band resonance. Indeed, the existence of a modulation in the electron density in 1L-$MoS_2$ has been demonstrated in previous studies on 1L-$MoS_2$ on a PPLN surface, revealing a photoinduced charging process in which light induces charge generation and transfer at the monolayer/substrate interface. This photodoping process strongly depends on the orientation ($P_{up}$ or $P_{down}$) of the ferroelectric domains.[24] In particular, photoluminescence experiments revealed that the downward band bending at the polar surface of the $P_{up}$ domains facilitates the accumulation of photoexcited electrons at the 1L-$MoS_2$/$LiNbO_3$ interface, leading to variations in the electron density of 1L-$MoS_2$ of around $5\times10^{14}$ $cm^{-2}$ under intense visible illumination. In contrast, on the $P_{down}$ domains, the electron density remains practically constant, regardless of the illumination intensity, highlighting the asymmetric role of the domain surfaces in governing carrier density modulation.[24]

To assess the impact of this domain-dependent photodoping process on the nonlinear optical response of the monolayer, we analyzed the evolution of the SHG modulation depth as a function of the fundamental beam power. As shown in **Figure 4**a, at low fundamental power, the modulation value of SHG is close to zero, indicating that the ferroelectric domain pattern has a negligible influence on the SHG response under these conditions. However, as the intensity of the fundamental beam increases, the contrast between the SHG response generated by the monolayer on the $P_{up}$ or $P_{down}$ domains increases markedly, consistent with the above-mentioned ferroelectric-domain-dependent photodoping process.[24] The corresponding SHG power curves for 1L-$MoS_2$ on each type of domain are shown in Figure 4b. As observed, the SHG intensity is systematically higher on the $P_{up}$ domain compared to that on the $P_{down}$ one, evidencing a light-driven, ferroelectric-domain-dependent SHG enhancement in the $P_{up}$ regions. The disparity between domains is further examined in the inset, which displays the data on a log–log scale. In the $P_{down}$ domain, the SHG response follows the ideal quadratic dependence (slope ≈ 2), in excellent agreement with the expected second-order process. Strikingly, the SHG signal on the $P_{up}$ domain exhibits a distinctly supra-quadratic behavior, with a fitted slope close to 2.3, pointing to an effective local enhancement of the $\chi^{(2)}$ value of 1L-$MoS_2$ on this domain orientation.

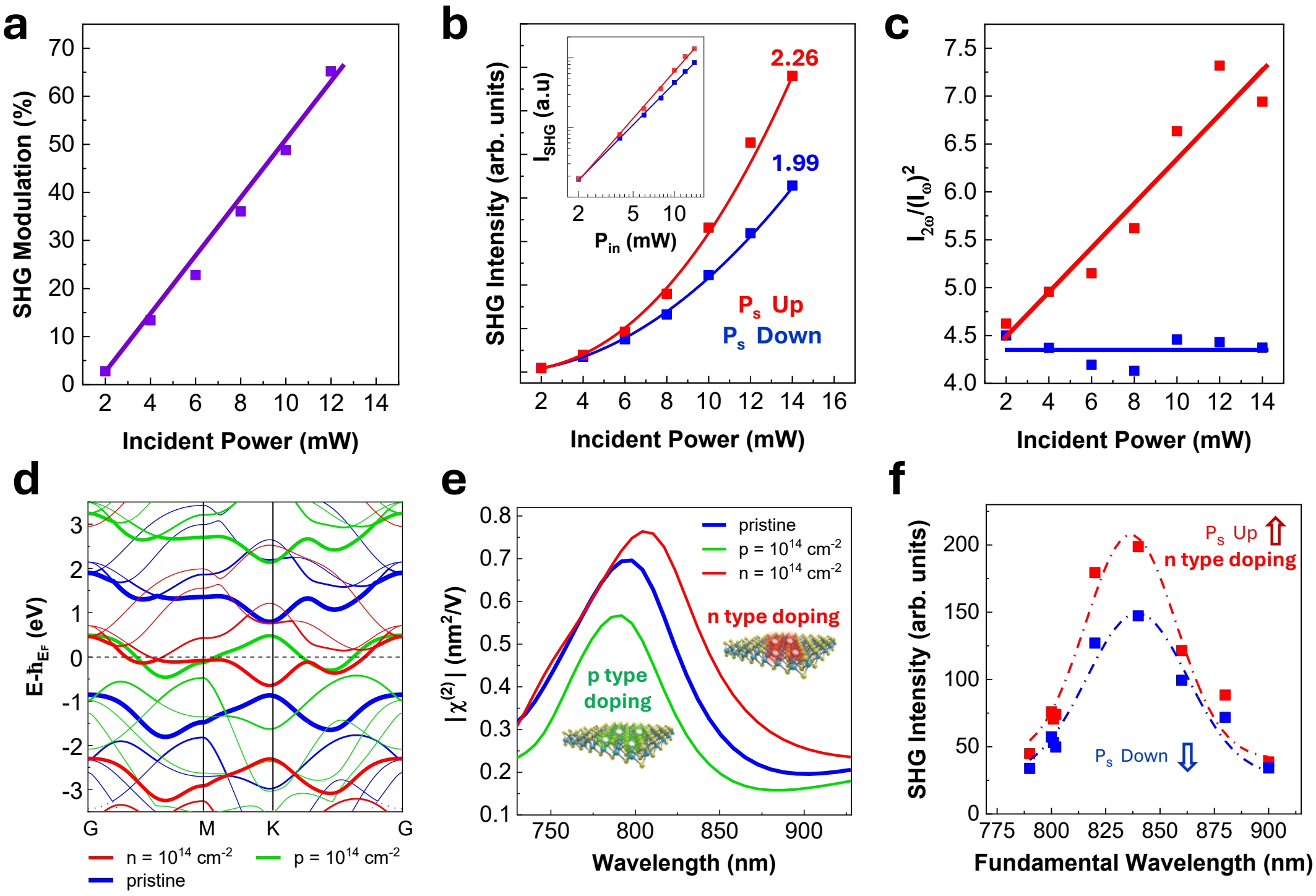

**Figure 4. Optically driven ferroelectric modulation of $\chi^{(2)}$.** a) Evolution of the SHG spatial modulation as a function of the incident power for the 1L-$MoS_2$/(z-cut) PPLN heterostructure. b) SHG power curves of 1L-$MoS_2$ on $P_{down}$ (blue) and $P_{up}$ (red) ferroelectric domains. The inset shows the output-power dependence in logarithmic scale along with the corresponding fittings (solid lines). c) Evolution of the SHG intensity normalized by the square of the fundamental power for $P_{down}$ (blue) and $P_{up}$ (red) domains, used to track relative variations in the effective $\chi^{(2)}$. Calculated d) electronic band structure of 1L-$MoS_2$, and e) $\chi^{(2)}$ values for pristine (blue), hole-doped (green), and electron-doped (red) cases. f) Fundamental-wavelength dependence of the SHG intensity of 1L-$MoS_2$ on $P_{down}$ (blue) and $P_{up}$ (red) ferroelectric domains. Dashed lines are a guide to the eye.

In fact, the difference in the effective $\chi^{(2)}$ between opposite domains can be estimated by normalizing the SHG intensity, $I_{2\omega}$, by the square of the fundamental intensity, $I_{\omega}$, taking to account the expression $I_{2\omega} \propto |\chi^{(2)}|^2 I_{\omega}^2$. Figure 4c depicts the incident-power dependence of the normalized SHG intensity $I_{2\omega}/I_{\omega}^2$ for each type of domain. On $P_{down}$ domains, this normalized quantity remains essentially constant over the explored power range, denoting a power-independent value of $\chi^{(2)}$. In contrast, on $P_{up}$ domains, the normalized intensity exhibits a clear increase with incident power, in agreement with a light-driven enhancement of $\chi^{(2)}$. The enhancement of the SHG response follows the same trend as the evolution of the electron density in 1L-$MoS_2$ on both $P_{up}$ and $P_{down}$ domains, as obtained from PL measurements (**Figure S4**). Specifically, a domain-dependent photodoping process leads to an increase in the electron density on the $P_{up}$ domain, while it remains essentially constant on the $P_{down}$ domain. The strong correlation between the SHG response and the photodoping inferred from PL measurements indicates that the observed SHG enhancement originates from photodoping-induced changes in the electronic density, which in turn enhance the nonlinear optical susceptibility of the monolayer on the $P_{up}$ regions.

In fact, the doping modulation is provided in a local and light-controlled manner by the ferroelectric substrate, without requiring electrical contacts or external gating. The electrode-free optical control of electron doping in the system arises from the combined effect of domain-dependent band bending and the resulting electric field at the polar surfaces, which facilitates the separation and transfer of photoexcited electron–hole pairs at the $MoS_2$-$LiNbO_3$ interface. At low excitation levels, few carriers are generated, and the photodoping effect remains minimal. However, carrier generation becomes more significant as the incident power increases, allowing the interfacial field to drive electrons to accumulate on $P_{up}$ domains, where downwards band bending becomes more prominent. This can give rise to variations in electron density of the order of $10^{14}$ $cm^{-2}$ as estimated by analyzing the spectral weight ratio of trions to excitons in the PL spectra. It also explains the constant value of $\chi^{(2)}$ observed in the $P_{down}$ domains, where the electronic density remains nearly invariant with increasing excitation light intensity. The results are in excellent agreement with those obtained from linear optics (Figure S4) and further highlight the key role of light excitation in modulating the electronic and optical responses in 2D–ferroelectric heterostructures.

To gain insight into the role of doping in the second-order nonlinear optical response, we employed a first-principles framework combined with a perturbative evaluation of the nonlinear optical response to calculate the electronic and SHG spectra of 1L-$MoS_2$ at varying charge carrier density. Our approach follows the implementation in the GPAW code, which applies the independent particle approximation (IPA).[44,45] The large doping levels considered in this study result in the filling of the conduction band (see band structures in Figure 4d). This effectively increases the metallic character of the material, thereby quenching the low-energy excitonic resonances (A and B) related to transitions at the K-point. While previous studies have shown that including many-body effects in the calculation of $\chi^{(2)}$ for undoped 1L-$MoS_2$ is crucial at excitonic resonances, the SHG resonance peak related to the C-interband transition (between the K- and Γ-points, where the conduction band remains empty even at high doping levels) is predicted only to exhibit a small blueshift and an increase of the intensity, while qualitative features remain similar to IPA predictions.[46]

Figure 4e shows a detailed view of the simulated second-order nonlinear susceptibility of 1L-$MoS_2$ as a function of n- and p-type doping levels, focusing on the C interband transition (two-photon resonance). Additional data over a wider spectral range can be found in **Figure S5**. The simulations support a clear doping dependence of $\chi^{(2)}$ around 800 nm with increasing intensity as the electronic charge carrier density increases. The effect is opposite for p-type doping, resulting in a decrease of $\chi^{(2)}$ rather than enhancement. Figure 4f compares the experimental SHG intensity recorded on domains with opposite polarity as a function of the fundamental wavelength within the studied spectral range. A clear enhancement of the SHG signal is observed on the $P_{up}$ domains, indicative of larger resonant values of $\chi^{(2)}$ under electron doping. Away from resonance, the SHG response of 1L-$MoS_2$ on opposite ferroelectric domains becomes similar, indicating a reduced sensitivity to domain-dependent doping, in agreement with the $\chi^{(2)}$ values predicted from theoretical calculations (Figure 4e). The consistency between experimental and theoretical results further supports the interpretation of a domain-dependent, light-driven tuning of the second-order nonlinear response in the $MoS_2$/$LiNbO_3$ heterostructure, establishing a new route for reconfigurable frequency conversion processes in hybrid 2D-ferroelectric platforms.

## 3. Conclusion

In summary, we demonstrate a light-driven, ferroelectric-domain-selective modulation of second-harmonic generation in monolayer $MoS_2$ integrated with a periodically poled $LiNbO_3$ substrate. Spatially resolved SHG measurements reveal a robust contrast between domains with opposite polarization, with an enhancement of up to ~70% on upward-polarized domains. This modulation strongly depends on excitation wavelength and optical power, reaching its maximum near the two-photon resonance associated with the C interband transition of $MoS_2$.

A systematic analysis of the SHG response as a function of excitation conditions, polarization configuration, and substrate orientation rules out interference or interfacial symmetry-coupling effects as the dominant mechanism. Instead, our analysis reveals that the observed behavior is explained by a ferroelectric-domain-dependent photodoping process at the $MoS_2/LiNbO_3$ interface. Light-induced charge generation and transfer selectively increase the electron density on upward-polarized domains, while the carrier density remains essentially unchanged on downward-polarized domains, leading to a domain-dependent enhancement of the effective second-order nonlinear susceptibility under resonant conditions. First-principles calculations support this interpretation by showing that electron doping significantly enhances the second-order susceptibility of monolayer $MoS_2$ near the C interband transition, in good qualitative agreement with the experimental results, whereas hole doping produces the opposite trend.

Overall, our results establish ferroelectric substrates as active and reconfigurable platforms for tuning nonlinear optical responses in integrated photonic systems through purely optical means, without requiring electrical contacts or external gating. The combination of ferroelectric domain engineering and light-induced doping enables spatially selective and dynamically tunable frequency conversion, paving the way for advances in classical and quantum nonlinear optics, integrated photonics, all-optical sensing, and neuromorphic and memory computing applications based on hybrid 2D–ferroelectric platforms.

## 4. Experimental Section and Methods

*Sample Preparation and Characterization*: $MoS_2$ flakes were mechanically exfoliated from a bulk crystal (HQ Graphene) and initially deposited onto a polydimethylsiloxane (PDMS) sheet. These flakes were subsequently transferred to either polar (Z-cut) or nonpolar (Y-cut) surfaces of a 0.5 mm thick commercial periodically poled lithium niobate crystals (Covesion Ltd) using a deterministic dry-transfer method.[47] To ensure adequate spatial resolution, crystals with domain periodicities of approximately 20 μm were employed. The layer number of $MoS_2$ flakes was identified using differential micro-reflectance spectroscopy across the visible spectral range. For these measurements, illumination from a halogen lamp was focused onto the sample through a 100× microscope objective (NA = 0.95), and the reflected signal was collected via an optical fiber coupled to a CCD spectrometer (Spectral Products CVI SM440). To further enhance the spatial resolution, a spatial pinhole (~1 μm) was introduced into the detection path.

*SHG Measurements:* SHG mapping was obtained by a customized laser-scanning confocal microscope (Olympus BX41) equipped with a motorized XY translation stage controlled via Labspec software. The positioning system offered a resolution of 0.3 μm. As excitation source, a tunable femtosecond Ti:sapphire laser (Spectra Physics, Model 177-Series), operating at a spectral range of 770 – 900 nm, was employed. The laser beam was focused onto the sample surface using a 50× (NA = 0.35) microscope objective. SHG emission was collected in backscattering configuration through the same objective and directed to a Horiba Synapse CCD connected to a Horiba iHR 550 monochromator.

*Computational Methods:* The ground-state electronic structure of monolayer $MoS_2$ was obtained using density functional theory (DFT) as implemented in the GPAW code.[44] A plane-wave basis was employed, with an energy cutoff of 600 eV as well as a k-point sampling of the Brillouin zone of (7,7,1) points to ensure convergence of the optical response. Doping was introduced by changing the total number of electrons in the unit and introducing a homogeneous compensating background. For each targeted electron (n-type) or hole (p-type) doping density, a separate self-consistent DFT calculation was performed.

The second-order nonlinear susceptibility $\chi^{(2)}$ was evaluated within the independent-particle approximation using a length-gauge perturbative formalism based on the obtained electronic ground state and the transition dipole matrix elements.[45] Both interband and intraband contributions were included by summing all possible transitions within the energy window of interest between the occupied and 10 lowest unoccupied bands. All spectra were broadened by a small phenomenological damping parameter of 50 meV to mimic finite lifetimes and facilitate comparison with realistic experimental linewidths.

## Acknowledgements

D.H-P, M.J F-M, J.D.C L.E.B, and M.O.R acknowledge funding from the Spanish State Research Agency MICIU/AEI/10.13039/501100011033 under grant PID2022-137444NB-I00 and “Maria de Maeztu” Programme for Units of Excellence in R&D CEX2023-001316-M. L.J. is supported by a career grant from the Novo Nordisk Foundation (grant no. NNF25OC0105410). G.L-P acknowledge funding from the Spanish State Research Agency under grant PID2022-138908NB-C32 and the Ramon y Cajal contract RYC2023-044003-I. J.D.C.is a Sapere Aude research leader supported by Independent Research Fund Denmark (grant no. 0165-00051B). The Center for Polariton-driven Light-Matter Interactions (POLIMA) is funded by the Danish National Research Foundation (Project No. DNRF165). Part of the computation in this project was performed on the DeiC Large Memory HPC system managed by the eScience Center at the University of Southern Denmark.

## Data Availability Statement

The data that support the findings of this study are available from the corresponding author upon reasonable request.

Supporting Information

***Domain-Selective Enhancement of Second Harmonic Generation in Monolayer $MoS_2$ via Ferroelectricity-Controlled Photodoping***

*David Hernández-Pinilla,[1,2] Line Jelver,[3] María Jesús Martínez-Morillo,[1,2] César Hernando-Fuente,[1] Miquel Cherta,[1] Guillermo López-Polin,[1,4] Joel D. Cox,[3,5] Luisa E. Bausá[1,2,4] and Mariola O Ramírez [1,2,4]**

1. *Dept. Física de Materiales, Universidad Autónoma de Madrid, Madrid 28049, Spain*
2. *Instituto Nicolás Cabrera, Universidad Autónoma de Madrid, Madrid 28049, Spain*
3. *POLIMA - Center for Polariton-driven Light−Matter Interactions, University of Southern Denmark, DK-5230 Odense M, Denmark*
4. *Condensed Matter Physics Center (IFIMAC), Universidad Autónoma de Madrid, Madrid 28049, Spain*
5. *Danish Institute for Advanced Study, University of Southern Denmark, DK-5230 Odense M, Denmark*

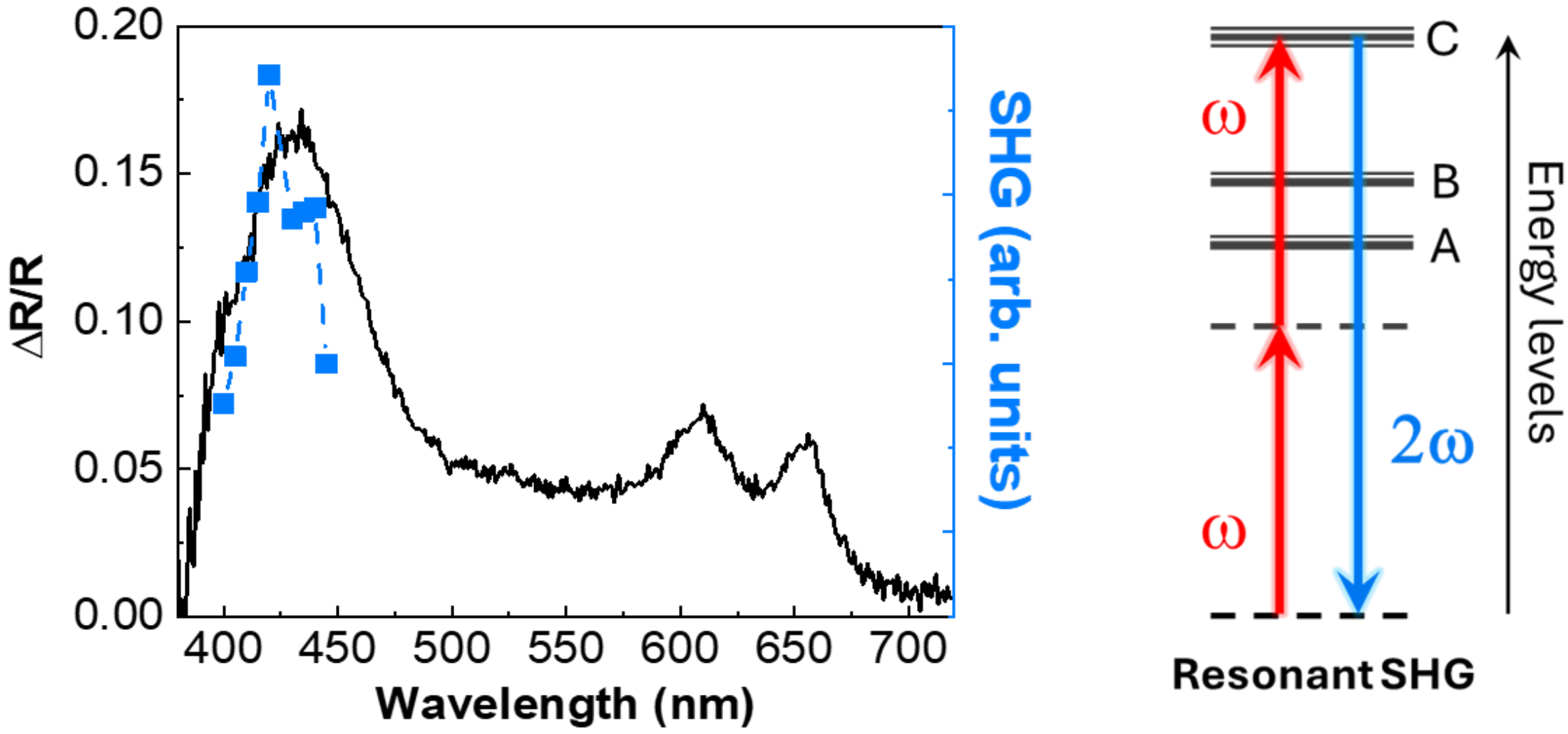

**Figure S1.** Spectral dependence of SHG intensity in 1L-$MoS_2$ (blue). The differential reflectance spectrum of 1L-$MoS_2$ (black) is shown for comparison with the C-band resonance. The right panel shows a schematic energy-level diagram for the two-photon resonant SHG process around the C-interband transition.

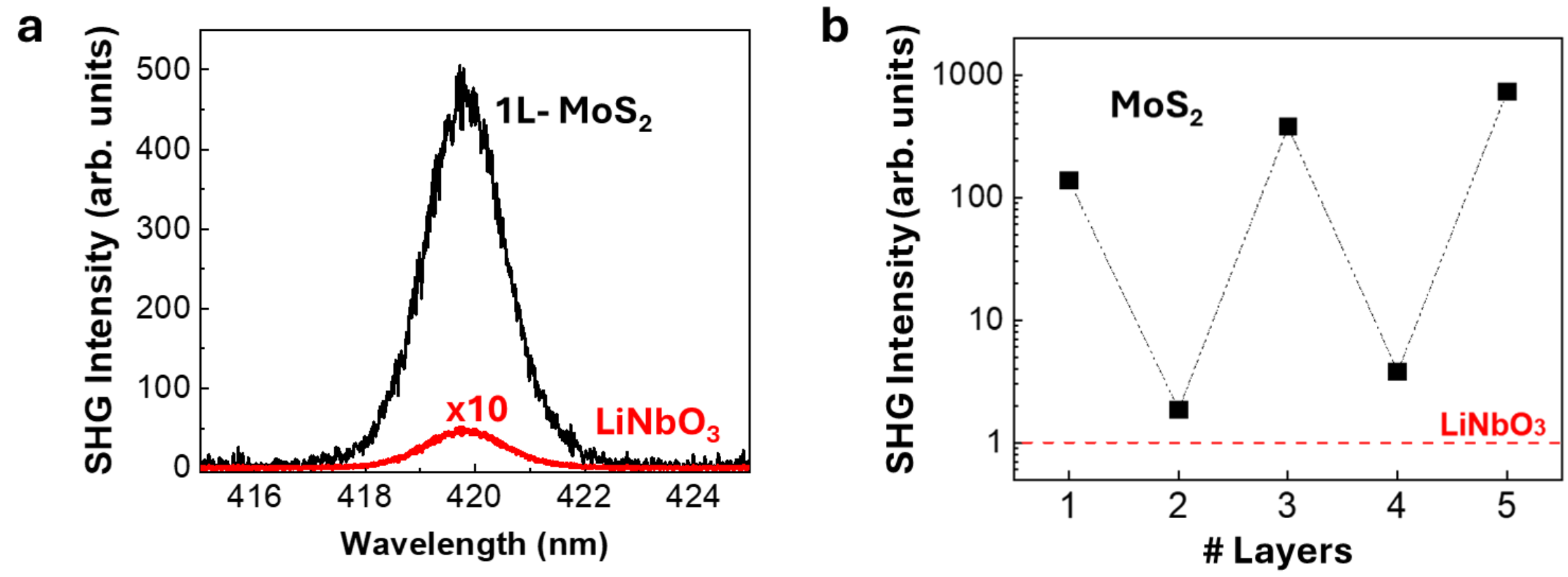

**Figure S2.** a) SHG spectra recorded from 1L-$MoS_2$ on $LiNbO_3$ (black) and from bare $LiNbO_3$ (red) for a fundamental wavelength of 840 nm. The $LiNbO_3$ signal is multiplied by a factor of 10. b) SHG intensity as a function of the number of $MoS_2$ layers (black), normalized to the corresponding SHG signal of bare $LiNbO_3$ (red dashed).

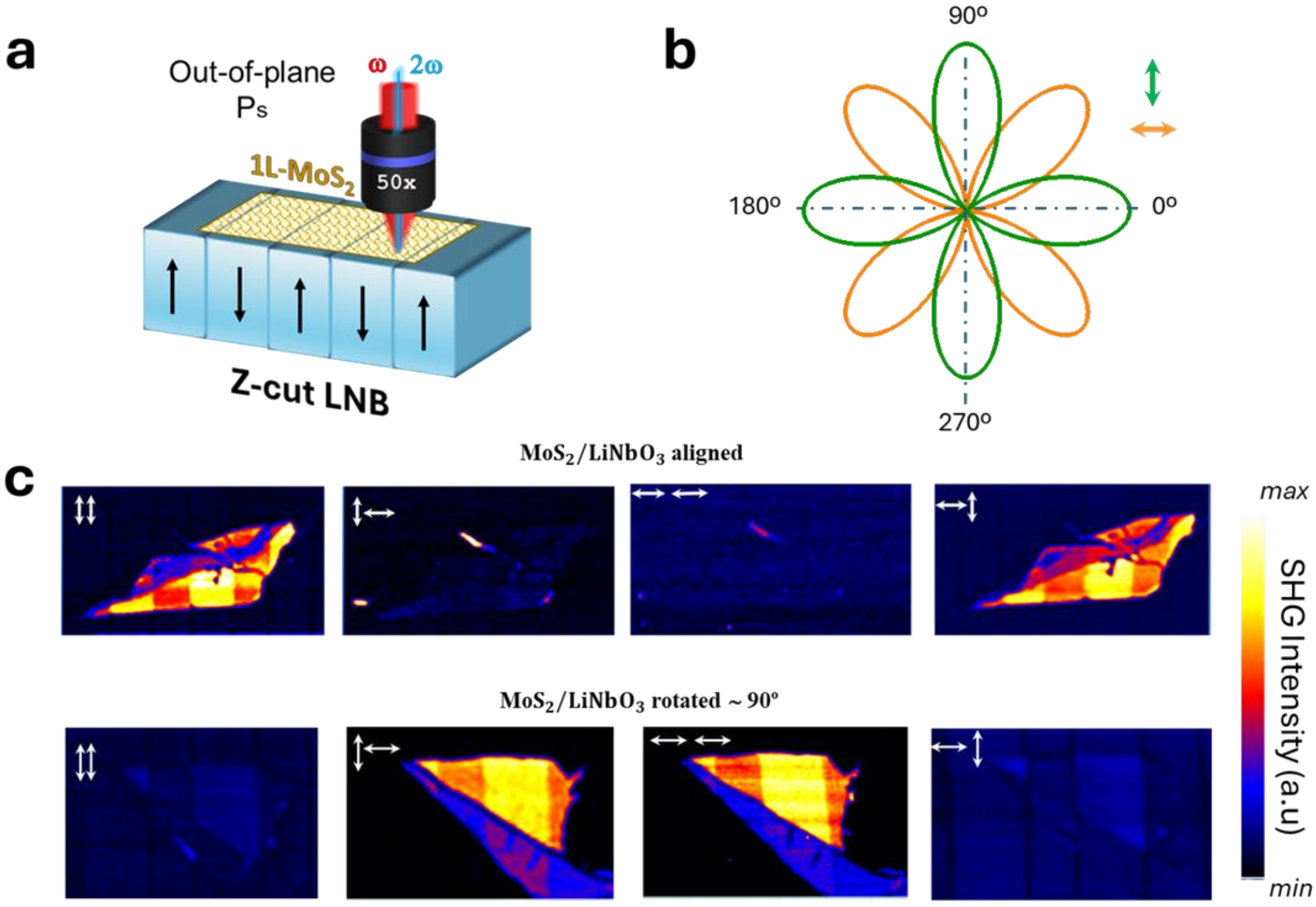

**Figure S3.** a) Schematic illustration of the 1L-$MoS_2$/$LiNbO_3$ heterostructure used in the SHG experiments. The spontaneous polarization ($P_s$) of $LiNbO_3$ is represented by black arrows indicating the out-of-plane (polar) configuration (Z-cut $LiNbO_3$). Fundamental beam (red) and second-harmonic signal (blue) collected in reflection geometry. b) Theoretical SHG response of 1L-$MoS_2$ as a function of the fundamental polarization when the fixed analyzer is parallel (green) or perpendicular (orange) to the domain walls. c) Spatial distribution of the integrated SHG intensity recorded at normal incidence for both orthogonal and parallel input-output polarization configurations in two different flakes, with the armchair direction of the monolayer oriented parallel (top) and perpendicular (bottom) to the ferroelectric domain walls of $LiNbO_3$. The two arrows in the maps indicate the linear polarization state of the fundamental and the corresponding SHG signal, respectively.

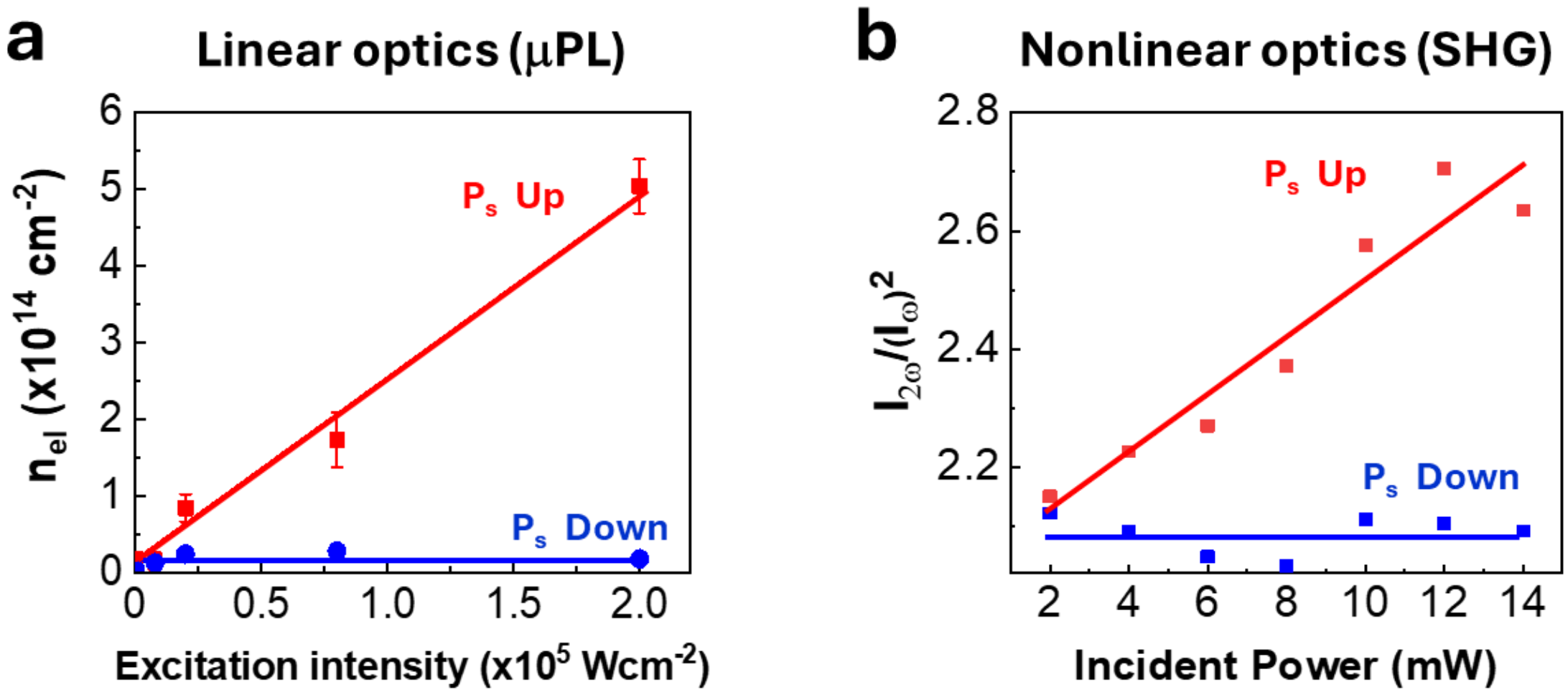

**Figure S4.** a) Electron density of 1L-$MoS_2$ /$LiNbO_3$ as a function of the excitation intensity, obtained from photoluminescence measurements in the $P_{up}$ (red) and $P_{down}$ (blue) domains. b) Evolution of the SHG intensity normalized by the square of the fundamental power as a function of incident power for $P_{down}$ (blue) and $P_{up}$ (red) domains. This representation allows tracking relative variations in the effective $\chi^{(2)}$ induced by electron doping.

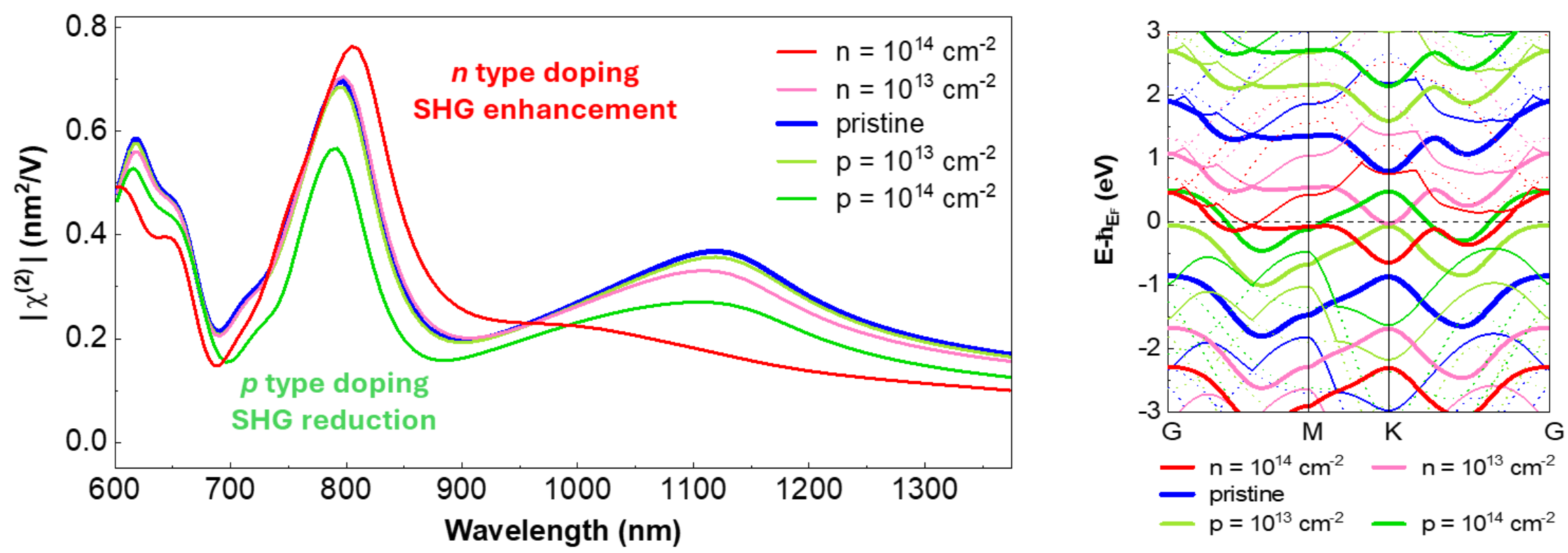

**Figure S5.** Calculated $\chi^{(2)}$ values (left panel) and electronic band structure of 1L-$MoS_2$ (right panel) for pristine (blue), and hole-doped (green and light green) and electron-doped (red and pink) cases corresponding to doping concentrations of $10^{14}$ $cm^{-2}$ and $10^{13}$ $cm^{-2}$, respectively.